\documentclass[conference]{assets/IEEEtran}
\IEEEoverridecommandlockouts
\usepackage{cite}

\usepackage{amsmath,amssymb,amsfonts}
\usepackage{algorithmic}
\usepackage{graphicx}
\usepackage{amsmath}
\usepackage[inline]{enumitem}
\usepackage{textcomp}
\usepackage{pgfplots}
\pgfplotsset{compat=1.18}
\usepackage{subcaption}
\usepackage{pgfplotstable}
\usepackage{dblfloatfix}
\usepackage{url}
\usepackage{soul}
\usepackage{hyperref}
\usepackage{xcolor}
\usepackage{subcaption}
\usepackage{lipsum}
\usepackage[inline]{enumitem}
\usepackage{enumitem}

\usetikzlibrary{arrows.meta, positioning}
\usetikzlibrary{shapes.geometric}
\usetikzlibrary{calc}
\usepgfplotslibrary{statistics}

\def\BibTeX{{\rm B\kern-.05em{\sc i\kern-.025em b}\kern-.08em
    T\kern-.1667em\lower.7ex\hbox{E}\kern-.125emX}}
\begin{document}

\title{Open Sky, Open Threats: Replay Attacks in Space Launch and Re-entry Phases\\
}

\author{\IEEEauthorblockN{Nesrine Benchoubane}
\IEEEauthorblockA{\textit{Polytechnique Montréal}\\
Montréal, QC, Canada \\
nesrine.benchoubane@polymtl.ca}
\and
\IEEEauthorblockN{Eray Güven}
\IEEEauthorblockA{\textit{Polytechnique Montréal}\\
Montréal, QC, Canada \\
guven.eray@polymtl.ca}
\and
\IEEEauthorblockN{Gunes Karabulut Kurt}
\IEEEauthorblockA{\textit{Polytechnique Montréal}\\
Montréal, QC, Canada \\
gunes.kurt@polymtl.ca}
}

\maketitle

\begin{abstract}
This paper examines the effects of replay attacks on the integrity of both uplink and downlink communications during critical phases of spacecraft communication. By combining software-defined radios (SDRs) with a real-time channel emulator, we replicate realistic attack conditions on the Orion spacecraft’s communication systems in both launch and reentry.  Our evaluation shows that, under replay attacks, the attacker’s signal can overpower legitimate transmissions, leading to a Signal to Noise Ratio (SNR) difference of up to $-7.8$ dB during reentry and $-6.5$ dB during launch. To mitigate these threats, we propose a more secure receiver design incorporating a phase-coherency-dependent decision-directed (DD) equalizer with a narrowed phase-locked loop (PLL) bandwidth. This configuration enhances resilience by making synchronization more sensitive to phase distortions caused by replay interference. 
\end{abstract}


\section{Introduction}

Artemis, led by NASA in collaboration with international partners including the European Space Agency (ESA), the Japan Aerospace Exploration Agency (JAXA), and the Canadian Space Agency (CSA), represents the next phase of human space exploration. The program aims to return astronauts to the lunar surface by the mid-2020s, establish a sustainable presence in cislunar space, and lay the groundwork for future crewed missions to Mars and beyond \cite{9172323}. A cornerstone of the Artemis mission architecture is the Orion spacecraft, engineered for high reliability, system redundancy, and operational resilience where it functions as the primary crew module \cite{8396769}.

\subsection{Orion Spacecraft}

Orion employs $\sim 6$ MHz bandwidth in S band communication system that utilizes four phased array antennas on the crew module and two on the service module, which are electronically steerable to enable dynamic beamforming for command uplink, telemetry downlink, and full-duplex voice and video transmission without mechanical repositioning \cite{nasa_orion}. 

These capabilities are central to both Artemis I and Artemis II, which form the foundation for testing Orion’s performance in an increasingly complex cislunar environment. Artemis I, the uncrewed precursor, launched aboard the Space Launch System (SLS) and proceeded through a sequence of high-dynamic events: booster jettison, service module panel separation, core stage shutdown, and translunar injection \cite{eckman2023trajectory}. Its return phase demonstrated a novel skip re-entry technique, where Orion dipped into the atmosphere, exited, then reentered for final descent—enhancing landing precision while reducing thermal loads \cite{rea2024orion}. 
% Fig.~\ref{fig:orion} illustrates Orion’s communication architecture in relation to key mission phases.

Two of the most vulnerable mission phases are launch and reentry, where system timing, communication integrity, and command execution are most critical. The Ascent Abort-2 test demonstrated Orion’s capability to autonomously trigger crew module separation during ascent in response to anomalies, a function that, if disrupted, could lead to mission failure or crew risk. Similarly, during reentry, any disruption could affect guidance, trajectory execution, or parachute deployment, leading to mission termination or failure to recover the spacecraft. Both phases, operating under critical timing constraints, are prime targets for adversarial interference, which can exploit vulnerabilities in command signals and system response sequences.

The wide beamwidth characteristics of Orion's tracking system help reduce pointing errors and mitigate the risk of communication outages. However, wider beam radiation inherently exposes the system to increased susceptibility to physical-layer threats. Additionally, while higher frequency operations offer greater bandwidth, they impose stricter root mean square (RMS) surface tolerance requirements on the antenna subsystem \cite{schaire2021analysis}. This constraint complicates the design of precise Pointing, Acquisition, and Tracking (PAT) mechanisms, making the adoption of broader radiation patterns a trade-off between robustness and physical-layer vulnerability

\subsection{Physical-Layer Vulnerabilities}

Replay attacks represent a critical and pervasive threat vector across the entire space system architecture, encompassing the space segment, ground segment, and the communication links between them \cite{bailey2021cybersecurity, ccsds3501g3_2022}. In this attack, legitimate transmissions—such as command sets sent from ground control to the spacecraft—can be intercepted, recorded, and maliciously retransmitted at a later time. If not properly authenticated or time-validated, these replayed commands would be accepted and executed a second time, potentially leading to unintended behaviors such as redundant maneuvers, or disruption of mission sequences \cite{Morioka2024}. Despite the presence of perimeter defenses, these attacks often target the physical layer, exploiting weaknesses in synchronization and command integrity protocols. The threat is exacerbated by the widespread availability of low-cost software-defined radios (SDRs), which enable adversaries to mount such attacks without the need for advanced or nation-state-grade infrastructure \cite{9256695, Estevez2022Decoding}. 

Recent studies further underscore the gravity of this threat. \cite{Lenhart_2021} has demonstrated the feasibility of coordinated replay attacks using two colluding adversaries to relay and spoof Global Navigation Satellite System (GNSS) signals over long distances, highlighting the vulnerability of receivers to time-shifted, but otherwise valid, transmissions.  Similarly, security analyses of the Galileo protocol have revealed structural weaknesses that leave even cryptographically authenticated signals vulnerable to replay under certain conditions \cite{wang2023replay}. However, these works largely focus on identifying vulnerabilities in already deployed systems rather than exploring phase-specific mission operations or modeling how critical communication windows can be deliberately exploited by adversaries.

\subsection{Contributions}

This work presents a notional radio frequency (RF) replay attack targeting space-ground communication during mission-critical phases—specifically, launch and reentry. Using the Orion spacecraft as a representative high-value asset, we construct a threat model where an adversary replays previously captured transmissions under realistic channel and timing conditions. To the best of our knowledge, this is the first emulation of physical-layer replay attacks in such scenarios using SDRs combined with a hardware-in-the-loop channel emulator. Our contributions include:
\begin{itemize}
     \item A novel identification and modeling of communication vulnerabilities in Orion's launch and reentry phases, demonstrating susceFptibility to RF replay attacks and observing significant signal degradation at critical attacker output gains ($–25$ dB for reentry and $–45$ dB for launch).
     \item A novel SDR-based emulation framework that demonstrates how adversaries can compromise link integrity during launch/reentry phases.
    \item A proposed energy-efficient, event-driven replay-resilient protocol tailored for constrained mission environments. The countermeasures applied during the evaluation show a strong performance, reducing bit error rate (BER) by up to 54.5\% in reentry and 89\% during launch, even when the attacker signal at one ground station was overpowering, highlighting the potential for improved communication reliability under attack conditions.
\end{itemize}

\section{System Model}

We focus on two operationally sensitive mission phases: launch and reentry, where Orion communicates with the Space Communications and Navigation (SCAN) infrastructure \cite{doi:10.2514/6.2014-1689, doi:10.2514/6.2008-3587}. The communication system includes both fixed ground stations and mobile recovery vessels, supporting uplink (UL) and downlink (DL) services for telemetry, crew voice, and command data.

\subsection{Threat Model}

The threat model assumes an adversary operating from a High-Altitude Long-Endurance (HALE) platform equipped with SDR to intercept and replay mission communications. HALE platforms are leveraged in this scenario due to two key characteristics: \begin{enumerate*}[label=(\roman*)] \item their plausible deployment as part of mission support activities, allowing inconspicuous proximity to the operational area, and \item their ability to maintain continuous line-of-sight (LOS) with both the capsule and ground infrastructure during these critical communication windows \end{enumerate*}. 

Unlike terrestrial adversaries, HALE platforms operate above LOS obstructions and can loiter for extended durations, enabling prolonged interception opportunities during mission-critical phases. Accordingly, we examine two representative HALE platforms: the \textbf{Northrop Grumman RQ-4 Global Hawk (RQ-4)} \cite{usaf_rq4_2024} and the \textbf{General Atomics MQ-9 Reaper (MQ-9)} \cite{usaf_mq9_2024}, both of which possess the required altitude, endurance, and payload capacity necessary. For both the adversary settings, Table~\ref{tab:hale_comparison} summarizes the key specifications.

\begin{table}[tpb]
\centering
\renewcommand{\arraystretch}{1.5} 
\caption{Comparative performance characteristics of HALE platforms.}
\label{tab:hale_comparison}
\begin{tabular}{|l|c|c|}
\hline
\textbf{Platform} & \textbf{MQ-9 Reaper} & \textbf{RQ-4 Global Hawk} \\
\hline
Manufacturer & General Atomics & Northrop Grumman \\
\hline
Max Altitude & 50,000 ft (15,240 m) & 60,000+ ft (18,288+ m) \\
\hline
Endurance & 27 hours &  34 hours \\
\hline
Cruise Speed & 240 KTAS (444 km/h) & 310 KTAS (574 km/h) \\
\hline
Operational Role & Tactical Surveillance / ISR & Strategic ISR \\
\hline
Range & ~1,150 nautical miles & 12,300 nautical miles \\
\hline
\end{tabular}
\end{table}

% s Sun Tzu wrote in The Art of War, "When we are near, we must make the enemy believe we are far away; when far away, we must make him believe we are near." Replay attacks embody this principle by manipulating the perceived temporal or spatial origin of commands to mislead the system at a critical moment.

% ... modeling of replay 

% - 

\subsection{Launch Phase}

\begin{figure}[tp]
    \centering
    \includegraphics[width=1\linewidth]{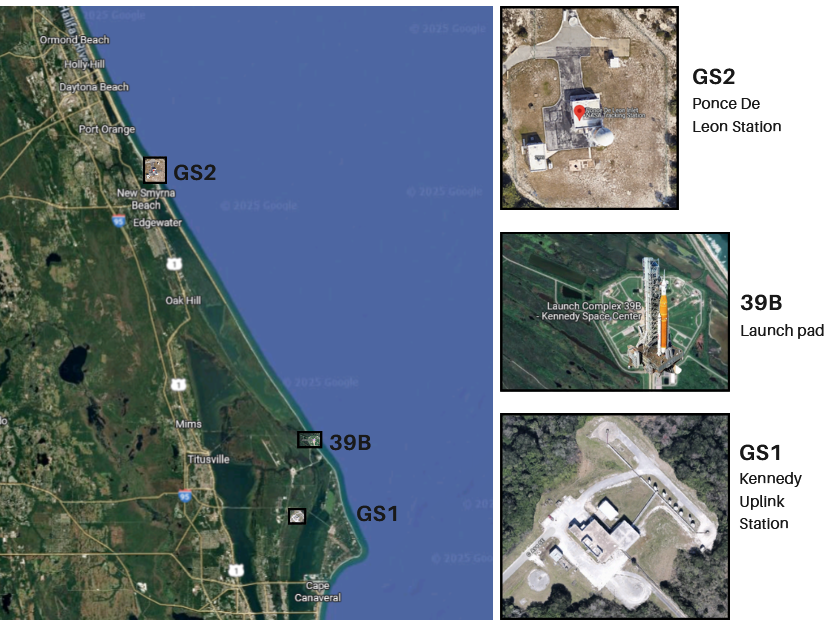}
    \caption{Illustration of the communication setup during launch, showing the relative positions of Kennedy Uplink Station (GS1) and Ponce De Leon Station (GS2) with respect to Launch Complex 39B.}
    \label{fig:ascent}
\end{figure}

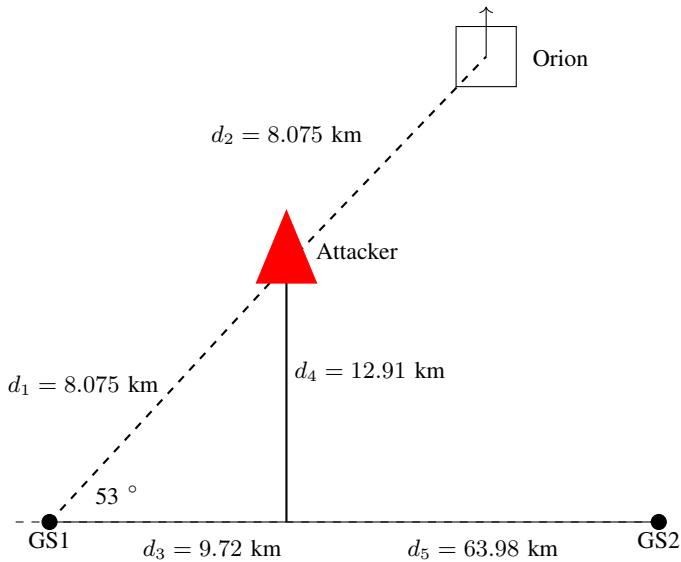
\begin{figure}[ht]
\begin{tikzpicture}[scale=0.9, every node/.style={font=\small}]
    % Base line and GS
    \coordinate (GS1) at (0,0);
    \coordinate (GS2) at (9,0);
    \coordinate (R1) at (4.8,0); 
    \coordinate (A) at (6.45,6.88); 
    \coordinate (att) at (3.5,3.8);
    \coordinate (gnd) at (3.5,0);
    \node[below] at (GS1) {GS1};
    \node[below] at (GS2) {GS2};
    \node[above right] at (7,6.6) {Orion};
    \node[right] at (3.8,4) {Attacker};

    \draw[dashed] (-0.5,0) -- (9,0);
    \node at (2.4,-0.4) {$d_{3} = 9.72$ km};
    \node at (6.4,-0.4) {$d_{5} = 63.98$ km};
    \node at (5.5,3,2) {$d_{4} = 12.91$ km};

    % Triangle and angle
    \draw[thick, dashed] (GS1) -- (att);
    \draw[thick, dashed] (att) -- (A);
    \draw[thick, black] (att) -- (gnd);
    \node[below left=5pt and 50pt, black, dashed] at (3.7,2.5) {$d_1 = 8.075$ km};
    \node[below left=5pt and 55pt, dashed] at (6.9,6.2) {$d_{2} = 8.075$ km};
    \node[black] at (1,0.4) {53 $^\circ$};

    \draw[black] (GS1) -- (GS2);

    \node[draw, fill=black, circle, minimum size=0.2cm, inner sep=0pt] at (GS1) {};

    \node[draw, fill=black, circle, minimum size=0.2cm, inner sep=0pt] at (GS2) {};

    \node[draw, rotate=90, minimum size=0.8cm] at (A) {};

    \node[isosceles triangle, fill=red, red, draw, minimum size=0.8cm, draw, rotate=90] at (att) {};

\draw[->, black] (6.45,6.88) -- (6.45,7.62);    \end{tikzpicture}
\caption{Geometric model illustrating the relative positions of the HALE platform, GS1, GS2, and Orion during the Max-Q phase of launch.}
\label{fig:geo-model-launch}
\end{figure}

During launch, Orion communicates with Earth through a primary ground segment composed of the Kennedy Uplink Station denoted GS1 and the Ponce De Leon Station denoted GS2. GS1, located approximately 9.72 km from Launch Complex 39B, is equipped with a 6.1-meter S-band antenna enclosed in a radome, while GS2 is situated an additional 63.98 km away from GS1, or about 56.45 km from the launch pad at 39B. This communication setup is shown in Fig.~\ref{fig:ascent}. These two stations support site diversity and are coordinated through a Best Frame Selector (BFS) mechanism, which dynamically selects the higher-quality signal between the two locations \cite{doi:10.2514/6.2014-1689}. 

In parallel, maintaining high-data-rate bi-directional links is essential for telemetry, command uplink, and crew safety \cite{NASA2017MILA}. A critical challenge to maintaining these links arises during the Max-Q period, when the vehicle experiences peak aerodynamic pressure. This phase, as demonstrated during Artemis I at an altitude of 12.97 km and speeds exceeding 469.3 m/s, places significant stress on the communication system under extreme dynamic conditions \cite{doi:10.2514/6.2008-3587}.

Thus, the focus of this study is the Max-Q phase, specifically around ~12.9 km altitude and ~467 m/s velocity, when both GS1 and GS2 maintain LOS with Orion. During this critical window, the adversary operating from the HALE platform between GS1 and Orion could intercept the DL communication. The geometric setup for this scenario is shown in Fig. \ref{fig:geo-model-launch}.

\subsection{Reentry Phase}

\begin{figure}[tpb]
    \centering
    \includegraphics[width=\linewidth]{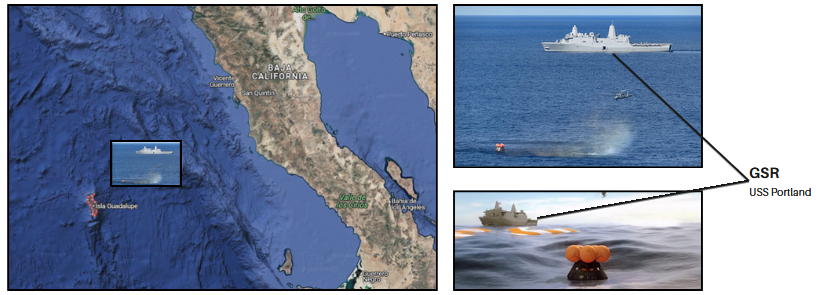}
    \caption{lustration of the communication setup during reentry showing the relative positions of Orion capsule and GSR positioned at sea.}
    \label{fig:descent}
\end{figure}

\begin{figure*}[tbp]
\centering
\begin{tikzpicture}[scale=1, every node/.style={font=\small}]
    % Coordinates
    \coordinate (GS) at (0,0);
    \coordinate (d3) at (0,5.62); 
    \coordinate (d4) at (5,5.62); 
    \coordinate (d_tau1) at (4,8.84); 
    \coordinate (L) at (12,0); 
    \coordinate (d1end) at (5.5,0); 
    
    \node[below] at (GS) {GSR};
    \node[above right=7pt] at (d4) {Orion};
    \node[above left=5pt] at (d3) {Attacker};

    \draw[thick] (GS) -- (d3) node[midway, left] {$d_2 = 6.7$ km};
    \draw[dashed] (d3) -- (d4) node[midway, below] {$d_3 = 5.62$ km};
    \draw[dashed] (GS) -- (d4) node[midway, right] {$d_{4} = 8.74$ km};
    \draw[thick] (GS) -- (L) node[midway, below] {$d_1 = 563$ km};
    \node[below] at (L) {Landing site};

    \node[draw, rotate=45, minimum size=0.8cm] at (d4) {};
    \node at ($(d4)+(-0.8,-0.5)$) {50$^\circ$};

    \node[isosceles triangle, fill=red, red, draw, minimum size=0.8cm, draw, rotate=90] at (d3) {};
    
    \node[draw, fill=black, circle, minimum size=0.2cm, inner sep=0pt] at (L) {};
    
    \node[draw, fill=black, circle, minimum size=0.2cm, inner sep=0pt] at (GS) {};

    \draw[->, black] (5,5.62) arc[start angle=60,end angle=0, radius=1];

\end{tikzpicture}
    \caption{Geometric model illustrating the relative positions of the HALE platform, GSR, and the descending capsule.}
    \label{fig:geo-model-descent}
\end{figure*}
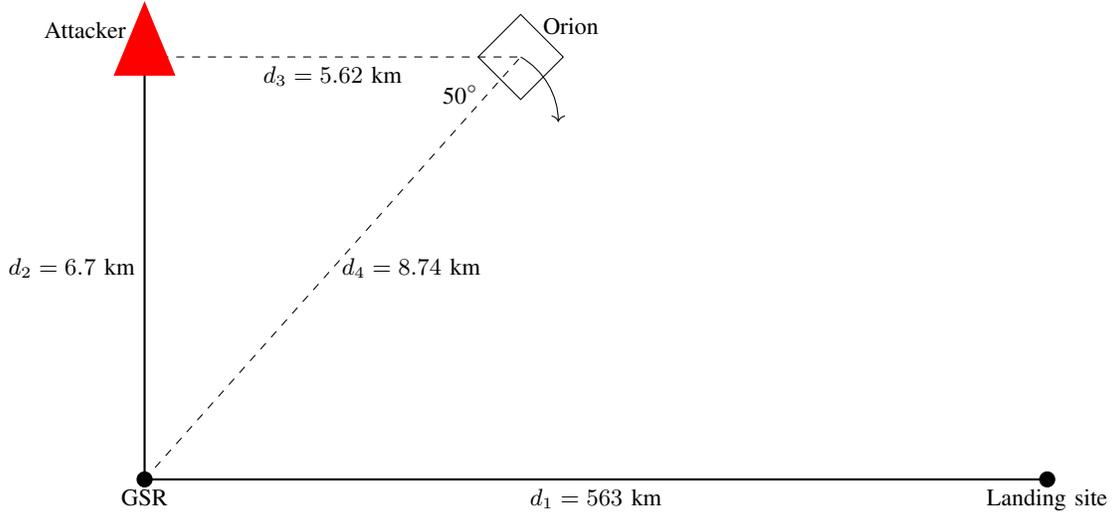

During reentry, Orion communicates with the SCAN network via contingency S-band links, with support from deployed Navy assets. The primary ground station (GSR) is a Navy amphibious recovery ship—such as the USS Portland—equipped with a well deck to enable recovery boats to dock with the capsule post-splashdown, stationed  $\sim 563$ km offshore near San Diego \cite{tingley2022navy}. This communication setup is shown in Fig.~\ref{fig:descent}. Both UL and DL channels are used during this phase to support contingency telemetry and voice communication between the Orion capsule and the GSR.

The reentry phase begins with high-velocity atmospheric entry, during which the spacecraft experiences an extended communications blackout caused by ionized plasma buildup around the heat shield. Communication is restored after this blackout, typically following the deployment of the drogue parachutes at an altitude of approximately 6.7 km and a descent speed of roughly 125 m/s.

At this drogue deployment altitude, the HALE platform situated above the recovery zone may attain LOS visibility with both the descending capsule and the GSR simultaneously. This introduces a critical vulnerability window in which both UL and DL communications are subject to interception. The geometric configuration of this scenario, depicting the relative positioning of the capsule, GSR, and the adversarial HALE platform, is shown in Fig.~\ref{fig:geo-model-descent}.

\section{Experimental Setup}

\begin{figure*}[ht]
\centering
    \includegraphics[width=\linewidth]{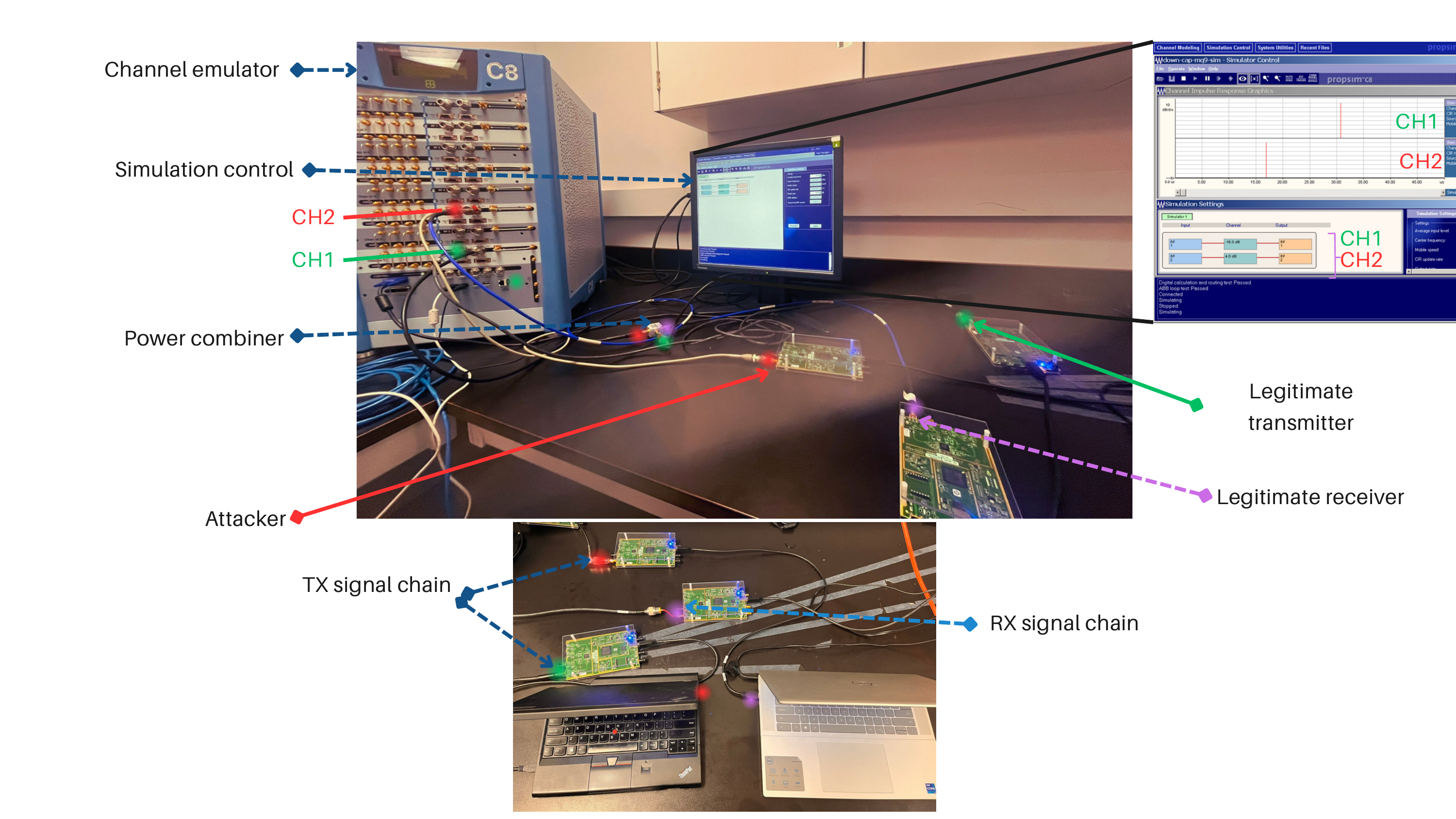}
    \caption{Overview of the experimental setup used to emulate replay attack scenarios during launch and reentry mission phases.}
    \label{fig:experimental setup}
\end{figure*}

The experimental setup for emulation consists of two key components: \begin{enumerate*}[label=(\roman*)] 
     \item a channel emulation stage, where realistic physical-layer conditions are simulated between communicating entities, and \item a device emulation stage, where SDRs are used to orchestrate transmissions and replay attacks
\end{enumerate*}. The goal is to replicate realistic replay attack scenarios under mission-relevant channel conditions for both UL and DL directions. Fig.~\ref{fig:experimental setup} illustrates the full experimental setup.

\subsection{Channel Emulation}
\label{channel-emulation}

We utilize the Electrobit Propsim C8 channel emulator to accurately reproduce wireless channel conditions between any two entities in the system, whether legitimate or adversarial. For each link, a two-tap multipath fading profile is instantiated, characterized by independent path delays, gains, and Doppler shifts. This allows for fine-grained emulation of both static and time-varying propagation effects.

The channel tap delays follow a sinusoidal model, where the delay oscillates around a mean delay with an amplitude for each sample. The Doppler spread is influenced by the angle between the incoming wave and the ground station, leading to time-variant channel fading. The channel impulse response (CIR) with $L$ multipaths can be described as:

\begin{equation}
    h(t,\tau) = \sum_{i=1}^{L} \beta_i(t) e^{j \phi_i(t)} \delta[\tau - \tau_i(t)] \hspace{1mm},
\end{equation}
where, $\beta_i(t)$ is the time varying amplitude of $i\textsuperscript{th}$ path, $\phi_i(t)$ is the phase of $i\textsuperscript{th}$ path in time of $t$ and $ \tau_i(t)$ is the delay corresponding to the $i\textsuperscript{th}$ path at time $t$. Either in UL or DL, the received signal is given by:
\begin{equation}
    y(t) = h(t,\tau) * x(t)  + n(t) \hspace{1mm},
\end{equation}
where $*$ denotes the convolution operator and $x(t)$ is the transmitted signal at carrier frequency of $f_c$. The transmitted signal is defined as:
\begin{equation}
    x(t) = I(t) \cos (2 \pi f_c t) - Q(t) \cos (2 \pi f_c t) \hspace{1mm},
\end{equation}
where $I(t)$ and $Q(t)$ are the Non-Return-to-Zero (NRZ) encoded in-phase and quadrature components of the message. Due to the mobility, each path is subject to Doppler shift of $\Delta f$ as following:
\begin{equation}
    \Delta f= \frac{\nu}{c}f_c \cdot \cos \alpha \hspace{1mm},
\end{equation}
where $\nu$ is the relative velocity between the two entities, $c$ is the speed of light, and $\alpha$ is the angle between mobile motion and incoming radio wave. In brief, $\Delta f$ is subject to estimation to be canceled as carrier frequency offset (CFO) in both GS1 and Orion. In this case, the received signal $y(t)$ with Doppler shift is expressed as:
\begin{align}
    y(t) = \sum_{i=1}^{L} \beta_i \bigg[ & I(t - \tau_i) \cos\left(2\pi (f_c + \Delta f_i) t + \phi_i(t) \right) \nonumber \\
      - &Q(t - \tau_i) \sin\left(2\pi (f_c + \Delta f_i) t + \phi_i(t)\right) \bigg] + n(t)
\end{align}
%As its clear, any residual Doppler shift estimation error directly impacts during the demodulation due to CFO.   

\subsection{Device Emulation and Replay Stages}
\label{replay stages}
Radio communication of the Orion, GS1, GS2, GSR, and HALE are demonstrated by USRP B200 model SDRs. Each wired connection is done by either SMA or Type-N connectors, which introduce insertion loss to the testbed. The setup consists of two stages:

\begin{enumerate}[label=\textbf{(Stage \arabic*)}, left=0pt, labelsep=0.4em]
\item \textbf{Exposure and Signal Capture}: The adversary eavesdrops the GS-Orion links over a legitimate (emulated) channel, recording the captured signal over the air. 
\item \textbf{Signal Replay}: The adversary replays the recorded signal toward the receiver, transmitting over a separate emulated channel. 
\end{enumerate}

The attacker is assumed to have access to soft information including the carrier frequency, trajectory, and spatiotemporal coordinates (position and time) of the target platform. This reflects a realistic replay threat model where partial physical-layer metadata exists. Additionally, due to independent propagation paths and lack of phase coherence, the attacker’s signal and the legitimate transmission are not coherently combined at the receiver, and thus the combining of the attacker and legitimate transmitter signals at the receiver is physically modeled through the channel emulator, preserving distinct multipath and Doppler characteristics for each path. 

Four phases of radio communication exist with the following definitions. The representation of the received signal during an attack is given in Equation \ref{pall}.

\begin{figure*}[tp]
\begin{equation}
    \label{pall}
y(t) = 
\left\{
\begin{aligned}
    &y_{GS1}(t) = h_{OR-GS1}(t, \tau) * x_{OR}(t) + n_{OR-GS1}(t) \\
    &\qquad + h_{ADV-GS1}(t, \tau) * y_{ADV}(t) + n_{ADV-GS1}(t), 
    &\hspace{3em} \text{Launch DL} \\
    &y_{GS2}(t) = h_{OR-GS2}(t, \tau) * x_{OR}(t) + n_{OR-GS2}(t), \\
    &\text{where } y_{ADV}(t) = h_{OR-ADV}(t, \tau) * x_{OR}(t) + n_{OR-ADV}(t)
    &\\[1em]
    &y_{OR}(t) = h_{GSR-OR}(t, \tau) * x_{GSR}(t) + n_{GSR-OR}(t) \\
    &\qquad + h_{ADV-OR}(t, \tau) * y_{ADV}(t) + n_{ADV-OR}(t), 
    &\hspace{3em} \text{Reentry UL} \\
    &\text{where } y_{ADV}(t) = h_{GSR-ADV}(t, \tau) * x_{GSR}(t) + n_{GSR-ADV}(t) &\\[1em]
    &y_{GSR}(t) = h_{OR-GSR}(t, \tau) * x_{OR}(t) + n_{OR-GSR}(t) \\
    &\qquad + h_{ADV-GSR}(t, \tau) * y_{ADV}(t) + n_{ADV-GSR}(t),
    &\hspace{3em} \text{Reentry DL} \\
    &\text{where } y_{ADV}(t) = h_{OR-ADV}(t, \tau) * x_{OR}(t) + n_{OR-ADV}(t)
\end{aligned}
\right.
\end{equation}
\end{figure*}

\subsubsection{Launch - DL}
Orion transmits to GS1 and GS2. The attacker eavesdrops on the transmission intended for GS1 and later replays it. Due to the physical separation of $d_5$, GS2 receives only the legitimate signal from Orion during the coverage. In contrast, GS1 captures a superposed signal consisting of the legitimate Orion transmission and the replayed DL attempt by the attacker.

\subsubsection{Reentry - UL}
GSR transmits to Orion while the attacker simultaneously replays a pre-recorded captured GSR signal.

\subsubsection{Reentry - DL}
Orion transmits to GSR, while the attacker replays the intercepted signal, causing GSR to receive a superposition of the legitimate and the replayed signal.

\subsection{Signal Processing Architecture}

A reciprocal DSP architecture is employed across both launch and reentry phases. Each DSP operation is applied to the complex baseband representation of the signal. The UL example during the reentry phase is detailed here. In this configuration, GSR consistently functions as the UL transceiver, while Orion serves as the transceiver. A similar architecture is adopted for the remaining analyzed signals, including: \begin{enumerate*}[label=(\roman*)]
     \item reentry DL, \item launch DL received at GS1, and \item launch DL received at GS2—where the latter is captured without attacker interference.
\end{enumerate*}

\subsubsection{Transceiver Architecture}

A predefined hexadecimal payload message $m_1$, consisting of $200$ bits, is transmitted to Orion. On GSR, firstly, the $m_1$ is upsampled, and the resulting pulse train is passed through a root raised cosine (RRC) pulse-shaping filter with a roll-off factor $\alpha = 0.35$. The baseband signal is generated at a sample rate of $250,000$ samples per second, and modulated with Quadrature Phase Shift Keying (QPSK) by a factor of four oversampling with NRZ encoding scheme. Resulting in the symbol rate of $62.500$ symbols per sec, the occupied bandwidth after RRC filtering is approximately $(1+\alpha) \times 62.500 = 84.375$ kHz. Following this, the baseband signal is completed and subsequently amplified by an on-chip, software-controllable AD9364 RF chip set to $53.6$ dB \cite{ettus_b200_series}. Lastly, amplified signal is then forwarded through the channel emulator to emulate the time-selective GSR-to-Orion channel. 

\subsubsection{Receiver Architecture }

The Orion node, implemented by a USRP B200 SDR, executes several stages to recover $m_1$. First, a differential detection stage filters the received signal over 32 uniformly spaced phases between $0$ and $2\pi$. Ideal sampling is performed with compensation for sampling clock offset between GSR and Orion. Prior to downsampling, the signal undergoes convolution with a matched filter with identical RRC taps used in transceiver to maximize SNR. This is followed by blind channel equalization using an adaptive FIR filter. Under no-attack conditions, amplitude variations remain low due to the LoS dominant channel with two significant taps, as described in Section~\ref{channel-emulation}. Hence, a direct decision adaptive filter is used as a linear equalizer. The receiver then estimates and corrects CFO using a digital phase-locked loop (DPPL) algorithm with a second-order loop bandwidth of $62.8 \times 10^{-3}$ \cite{li2000introduction} and concludes with demodulation. 

\subsubsection{Uplink - Attacker Architecture}

Operating under LOS but from a concealed position, the attacker records the transmission corresponding to $m_1$ at time $t_1$. The replay attack is carried out later at time $t_2$ by retransmitting the recorded signal in an effort to impersonate GSR.

\section{Assessment of Replay Attack Impact}

We conduct three experiments to evaluate the system's susceptibility to replay attacks under different signal conditions. In the first experiment, we vary the attacker’s output gain ($G_A^{out}$) during Stage 1 (as defined in Section \ref{replay stages}) to assess the maximum impact of signal injection during the exposure phase. In the second experiment, we fix $G_A^{out}$ and vary its receiver input gain ($G_A^{in}$) during Stage 2 to evaluate replay performance based on degraded or incomplete captures. In the third experiment, we vary the input gain of the legitimate transmitter during Stage 2 to assess the effects of an overpowering replay attack relative to a weakened legitimate signal.

Across all experiments, we report three metrics: BER, $\Delta \text{SNR}$, and received signal power levels. Referring to the SINR as SNR during the adversary attack, we denote as follows 
\begin{equation}
    \Delta {\textrm{SNR}} = \underbrace{\textrm{SINR [dB]}}_{\textrm{Under attack}} - \underbrace{\textrm{SNR [dB]}}_{\textrm{Under no-attack}}
\end{equation}
For BER, we calculate the ratio of erroneous bits to the total number of transmitted bits. The signal power refers to the digital power level (DPL), calculated by the downconverted IQ of $x[n]$ using noise averaging among $M$ samples, consistently applied throughout all experiments: 
\begin{equation}
\textrm{DPL} = \frac{1}{M} \sum^M \mathbb{R}(x[n])^2 + \mathbb{I}(x[n])^2 \hspace{1mm}. 
\end{equation}
Power levels are compared to the baseline to illustrate variations under different attack scenarios. Results are presented separately for each mission phase and communication direction.

\subsection{Reentry - DL}

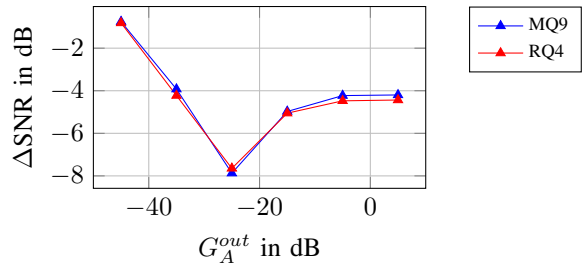
\begin{figure}[tp]
\raggedright
    \begin{tikzpicture}
    \begin{axis}[
        ylabel={$\Delta \text{SNR}$ in dB},
        xlabel={$G_A^{out}$ in dB},
        ylabel near ticks,
        % ymode=log,
        % log basis y=10,
        % ymin=1, ymax=20000,
        % xmin=0, xmax=550,
        % xtick={50, 100, 200, 500},
        % ytick={1, 10, 100, 1000, 10000},
        grid=both,
        height=4cm,
        width=6cm,
        legend style={at={(1.3, 1)}, anchor=north, legend cell align=left, font=\scriptsize}, 
    ]

    \addplot[color=blue, mark=triangle*] coordinates {(-45, -0.755843) (-35, -3.923924) (-25, -7.872367) (-15, -4.973906) (-5, -4.227751) (5, -4.197567)};
    \addlegendentry{MQ9}

    \addplot[color=red, mark=triangle*] coordinates {(-45, -0.816491) (-35, -4.226196) (-25, -7.647040) (-15, -5.054074) (-5, -4.473781) (5, -4.435433)};
    \addlegendentry{RQ4}
    \end{axis}
\end{tikzpicture}
\caption{$\Delta \text{SNR}$ under varying $G_A^{out}$ during Reentry-DL.}
\label{fig:dl-reentry-delta-snr}
\end{figure}

\begin{table}[tp]
\renewcommand{\arraystretch}{1.3} 
\centering
\caption{BER under varying $G_A^{out}$ during Reentry-DL.}
\label{tab:dl-reentry-ber-exp}

\begin{tabular}{c>{\centering\arraybackslash}p{2cm}cccc}
\hline
\textbf{Type} & \textbf{$G_A^{in}$ (Stage 2) in dB} & \multicolumn{3}{c}{\textbf{$G_A^{out}$ (Stage 1) in dB}} \\
\cline{3-5}
 & &  \textbf{-25} & \textbf{-35} & \textbf{-45} \\
\hline
\textbf{RQ4} & \textbf{-8}  & 10.5995 & 0.0003  & 0 \\
  & \textbf{-15}  & 50.365 & 0.1175 & 0.001 \\
\hline
\textbf{MQ9}  & \textbf{-8}  & 13.0047 & 0.0002 & 0.0008 \\
 & \textbf{-15}  & 50.8657 & 0.1242 & 0 \\
\hline
\end{tabular}
\end{table}

\begin{figure}[tp]
\raggedright
\begin{tikzpicture}
\begin{axis}[
height=5cm,
width=6cm,
grid=both,
ytick={2,4,6,7},
yticklabels={$-25$, $-35$, $-45$, Ref},
xlabel={Digital power level},
ylabel={$G_A^{out}$ in dB},
% boxplot/whisker style={solid},
boxplot/every median/.style={solid, thick},
boxplot/every box/.style={solid},
boxplot/every whisker/.style={solid},
boxplot/every outlier/.style={mark=*, draw opacity=1},
boxplot/every average/.style={solid},
]
% MQ9 -25 with minus8propsim
\addplot+ [
  boxplot prepared={
    lower whisker=60.000671,
    lower quartile=73.51412225,
    median=77.904945,
    upper quartile=82.747246,
    upper whisker=89.999977,
  },
  color=blue,
] coordinates {};
% RQ4 -25 with minus8propsim
\addplot+ [
  boxplot prepared={
    lower whisker=60.728691,
    lower quartile=73.62523825,
    median=77.813305,
    upper quartile=82.223015,
    upper whisker=89.999977,
  },
  color=red,
] coordinates {(0,60.000347) (0,60.001217) (0,60.00164)};

% MQ9 -35 with minus8propsim
\addplot+ [
  boxplot prepared={
    lower whisker=70.515182,
    lower quartile=73.61595725,
    median=74.4372025,
    upper quartile=75.683313,
    upper whisker=78.784317,
  },
  color=blue,
] coordinates {(0,60.540741) (0,61.378487) (0,61.492352)};

% RQ4 -35 with minus8propsim
\addplot+ [
  boxplot prepared={
    lower whisker=68.178909,
    lower quartile=71.900368,
    median=72.847382,
    upper quartile=74.381355,
    upper whisker=78.10276,
  },
  color=red,
] coordinates {(0,60.74403) (0,62.094131) (0,62.691795)};

% MQ9 -45 with minus8propsim
\addplot+ [
  boxplot prepared={
    lower whisker=71.006592,
    lower quartile=72.126656,
    median=72.499252,
    upper quartile=72.873398,
    upper whisker=73.993378,
  },
  color=blue,
] coordinates {(0,60.019569) (0,60.826214) (0,61.370884)};

% RQ4 -45 with minus8propsim
\addplot+ [
  boxplot prepared={
    lower whisker=68.739883,
    lower quartile=70.037918,
    median=70.456345,
    upper quartile=70.903282,
    upper whisker=72.201309,
  },
  color=red,
] coordinates {(0,60.16888) (0,60.899715) (0,61.55217)};

\addplot+ [
  boxplot prepared={
    lower whisker=81.201767,
    lower quartile=82.278488,
    median=82.613396,
    upper quartile=82.99633,
    upper whisker=84.07309,
  },
  color=black,
] coordinates {};
\end{axis}
\end{tikzpicture}
\caption{DPL boxplot between reference and superimposed signal during reentry DL with $G_A^{in}$ in Stage 2 at $-8$ dB.}
\label{fig:dl-reentry-power}
\end{figure}
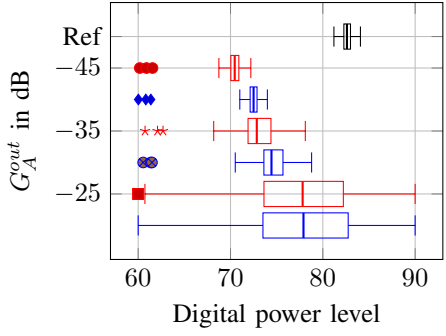

As shown in Fig.~\ref{fig:dl-reentry-delta-snr}, the $\Delta \text{SNR}$ remains minimal at low attacker output levels (e.g., $–0.76$ dB for MQ-9 and $–0.82$ dB for RQ-4 at $–45$ dB $G_A^{out}$), but begins to degrade significantly from $–25$ dB onwards, reaching roughly $–7.8$ dB. This indicates that at higher replay power, the attacker begins to dominate the channel, producing effects similar to destructive interference or overpowering of the legitimate signal. BER values in Table~\ref{tab:dl-reentry-ber-exp} support this: at –$25$ dB output gain, BER jumps to $13.0\%$ (MQ-9) and $10.6\%$ (RQ-4), while remaining negligible ($<0.001$\%) at –$35$ dB and below. Even when the attacker’s input gain is reduced to $–15$ dB, a strong enough output still results in BER reaching $50\%$, confirming the replay signal's capability to corrupt reception under realistic conditions.

\begin{table}[tp]
\renewcommand{\arraystretch}{1.3} 
\centering
\caption{$\Delta \text{SNR}$ under varying gains during reentry DL.}
\label{tab:dl-reentry-delta-snr-exp2}
\begin{tabular}{c>{\centering\arraybackslash}p{2cm}cccc}
\hline
\textbf{Type} & \textbf{$G_A^{in}$ (Stage 2) in dB} & \multicolumn{3}{c}{\textbf{$G_A^{out}$ (Stage 1) in dB}} \\
\cline{3-5}
 & &  \textbf{-25} & \textbf{-35} & \textbf{-45} \\
\hline
\textbf{RQ4} & \textbf{-8}  & -5.886978 & -1.345006 & -0.248249 \\
  & \textbf{-15}  &-7.647040 & -4.226196 & -0.816491 \\
\hline
\textbf{MQ9}  & \textbf{-8}  & -5.476275 & -1.274062 & -0.199777 \\
 & \textbf{-15} & -7.872367 & -3.923924 & -0.755843 \\
\hline
\end{tabular}
\end{table}

\begin{table}[h!]
\renewcommand{\arraystretch}{1.3} 
\centering
\caption{$\Delta \text{SNR}$ under varying $G_O^{in}$ Stage 2 during reentry DL.}
\label{tab:reentry-dl-overpowering}
\begin{tabular}{c>{\centering\arraybackslash}p{2cm}c}
\hline
\textbf{Type} & \textbf{$G_O^{in}$ (Stage 2) in dB}& {$\Delta \text{SNR}$ in dB} \\
\hline
\textbf{RQ4} &  \textbf{-15}    &-7.647040  \\
 & \textbf{-10}    & -5.944793  \\
 & \textbf{-5}    & -3.417873 \\
 & \textbf{0}   & -1.901100  \\
\hline
\textbf{MQ9} & \textbf{-15}  & -7.872367  \\
 & \textbf{-10}   & -6.012164   \\
 & \textbf{-5} & -3.409762  \\
 & \textbf{0}  & -2.169645  \\
\hline
\end{tabular}
\end{table}

To further analyze replay attack performance, we fix the attacker’s output gain (Stage 1) and vary the input gain (Stage 2) to observe the effect on $\Delta \textrm{SNR}$ and BER. Fig.~\ref{fig:dl-reentry-power} presents boxplots of digital power levels at GSR for both MQ-9 and RQ-4 platforms in the reentry DL scenario, compared against the reference (no attack) signal. 

At higher $G_A^{out}$ ($–25$ dB) and input gain of $–8$ dB, the superposed signal shows considerable spread and elevated power levels, with broader interquartile ranges and increased outliers, indicating strong variation and overlap with the attacker's replayed signal. As $G_A^{out}$ decreases to $–35$ dB and $–45$ dB, the power distributions begin to tighten and shift downward, revealing that the attack becomes less effective and increasingly resembles the clean reference signal. This trend is confirmed by $\Delta \text{SNR}$ values reported in Table~\ref{tab:dl-reentry-delta-snr-exp2}, where higher input gain leads to stronger interference (e.g., $–5.47$ dB for MQ-9 and $–5.88$ dB for RQ-4 at $–8$ dB input gain and $–25$ dB output gain). Reducing the input gain weakens the attacker signal, decreasing $\Delta \text{SNR}$.

Finally, we assess the impact of increasing the Orion input gain ($G_O^{in}$) while keeping the $G_A^{out}$ fixed at $–25$ dB, a setting previously shown to significantly degrade signal quality. As shown in Table~\ref{tab:reentry-dl-overpowering}, increasing $G_O^{in}$ from $–15$ dB to $0$ dB progressively improves the $\Delta \text{SNR}$. But while higher legitimate gain mitigates the attack’s impact to some extent, the signal remains degraded, confirming that the attacker at $–25$ dB maintains substantial influence even under overpowering conditions. This configuration is subsequently used to evaluate the effectiveness of countermeasures.

\subsection{Reentry - UL}

\begin{figure}[tp]
\raggedright
    \begin{tikzpicture}
    \begin{axis}[
        ylabel={$\Delta \text{SNR}$ in dB},
        xlabel={$G_A^{out}$ in dB},
        ylabel near ticks,
        % ymode=log,
        % log basis y=10,
        % ymin=1, ymax=20000,
        % xmin=0, xmax=550,
        % xtick={50, 100, 200, 500},
        % ytick={1, 10, 100, 1000, 10000},
        grid=both,
        height=4cm,
        width=6cm,
        legend style={at={(1.3, 1)}, anchor=north, legend cell align=left, font=\scriptsize}, 
    ]

    \addplot[color=blue, mark=triangle*] coordinates {(-45, -0.006831) (-35, -3.121910) (-25, -7.110577) (-15, -4.526478) (-5, -6.611560) (5, -6.110918)};
    \addlegendentry{MQ9}

    \addplot[color=red, mark=triangle*] coordinates {(-45, -0.046018) (-35, -3.097965) (-25, -6.918450) (-15, -4.309865) (-5, -3.662071) (5, -3.608301)};
    \addlegendentry{RQ4}
\end{axis}
\end{tikzpicture}
\caption{$\Delta \text{SNR}$ under varying $G_A^{out}$ during reentry UL.}
\label{fig:ul-reentry-delta-snr}
\end{figure}
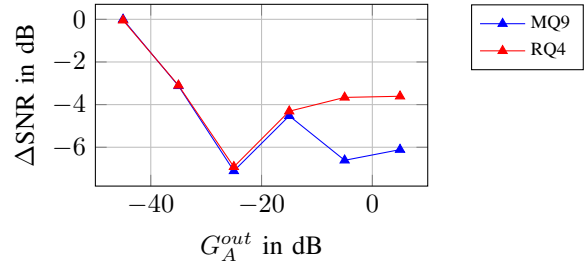

\begin{table}[tp]
\renewcommand{\arraystretch}{1.3} 
\centering
\caption{BER under varying $G_A^{out}$ during reentry UL.}
\label{tab:ul-reentry-ber-exp}
\begin{tabular}{c>{\centering\arraybackslash}p{2cm}cccc}
\hline
\textbf{Type} & \textbf{$G_A^{in}$ (Stage 2) in dB} & \multicolumn{3}{c}{\textbf{$G_A^{out}$ (Stage 1) in dB}} \\
\cline{3-5}
 & &  \textbf{-25} & \textbf{-35} & \textbf{-45} \\
\hline
\textbf{RQ4} & \textbf{-8}  &  30.2789 & 0.0002  & 0.0005 \\
  & \textbf{-15}  & 49.9606 & 0.0317 & 0 \\
\hline
\textbf{MQ9}  & \textbf{-8}  & 26.7524 & 0.0004 & 0 \\
 & \textbf{-15}  & 51.8184 & 0.0441 & 0 \\
\hline
\end{tabular}
\end{table}

A similar trend of reentry DL is observed in the UL case. As the $G_A^{out}$ exceeds $–25$ dB, the replay signal shifts from benign to highly disruptive, effectively over taking the UL. BER peaks at $30\%$ for RQ-4 and $26\%$ for MQ-9 at $–25$ dB, compared to near-zero levels at $-35$ dB and below (see Table~\ref{tab:ul-reentry-ber-exp}). Interestingly, the $\Delta \text{SNR}$ evolution differs from the DL: MQ-9 experiences a steep degradation up to $–7.1$ dB at –$25$ dB gain and sees partial recovery at higher gains (see Fig.~\ref{fig:ul-reentry-delta-snr}). Overall, we can see in the UL platform-dependent sensitivity in replay-induced signal degradation.

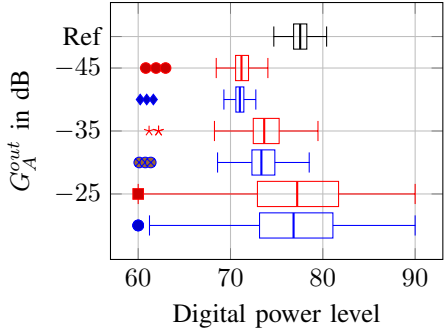
\begin{figure}[tp]
\raggedright
\begin{tikzpicture}
\begin{axis}[
height=5cm,
width=6cm,
grid=both,
ytick={2,4,6,7},
yticklabels={$-25$, $-35$, $-45$, Ref},
xlabel={Digital power level},
ylabel={$G_A^{out}$ in dB},
% boxplot/whisker style={solid},
boxplot/every median/.style={solid, thick},
boxplot/every box/.style={solid},
boxplot/every whisker/.style={solid},
boxplot/every outlier/.style={mark=*, draw opacity=1},
boxplot/every average/.style={solid},
]
% MQ9 -25 with uplink (minus8inputpropsim)
\addplot+ [
  boxplot prepared={
    lower whisker=61.224693,
    lower quartile=73.145048,
    median=76.831146,
    upper quartile=81.092165,
    upper whisker=89.999886,
  },
  color=blue,
] coordinates {(0,60.001541) (0,60.001545) (0,60.002853)};

% RQ4 -25 with uplink (minus8inputpropsim)
\addplot+ [
  boxplot prepared={
    lower whisker=60.000542,
    lower quartile=72.9170665,
    median=77.222355,
    upper quartile=81.704540,
    upper whisker=89.999939,
  },
  color=red,
] coordinates {(0,60.000542)};

% MQ9 -35 with uplink (minus8inputpropsim)
\addplot+ [
  boxplot prepared={
    lower whisker=68.598747,
    lower quartile=72.320969,
    median=73.3458215,
    upper quartile=74.80295775,
    upper whisker=78.525879,
  },
  color=blue,
] coordinates {(0,60.114452) (0,60.770504) (0,61.363178)};

% RQ4 -35 with uplink (minus8inputpropsim)
\addplot+ [
  boxplot prepared={
    lower whisker=68.262817,
    lower quartile=72.46688625,
    median=73.649273,
    upper quartile=75.26993175,
    upper whisker=79.474487,
  },
  color=red,
] coordinates {(0,61.17992) (0,62.202568) (0,62.21167)};

% MQ9 -45 with uplink (minus8inputpropsim)
\addplot+ [
  boxplot prepared={
    lower whisker=69.28466,
    lower quartile=70.581404,
    median=71.014061,
    upper quartile=71.445923,
    upper whisker=72.742699,
  },
  color=blue,
] coordinates {(0,60.222794) (0,60.924152) (0,61.644222)};

% RQ4 -45 with uplink (minus8inputpropsim)
\addplot+ [
  boxplot prepared={
    lower whisker=68.455147,
    lower quartile=70.55265,
    median=71.188263,
    upper quartile=71.951103,
    upper whisker=74.048744,
  },
  color=red,
] coordinates {(0,60.822208) (0,61.932709) (0,62.969257)};

% Ground truth - power_uplink
\addplot+ [
  boxplot prepared={
    lower whisker=74.69162,
    lower quartile=76.836151,
    median=77.547386,
    upper quartile=78.266693,
    upper whisker=80.410973,
  },
  color=black,
] coordinates {};

\end{axis}

\end{tikzpicture}
\caption{DPL level boxplot between reference and superimposed signal during reentry UL with $G_A^{in}$ in Stage 2 at $-8$ dB.}
\label{fig:ul-reentry-power}
\end{figure}

\begin{table}[tp]
\renewcommand{\arraystretch}{1.3} 
\centering
\caption{$\Delta \text{SNR}$ under varying gains during reentry UL.}
\label{tab:ul-reentry-delta-snr-exp2}
\begin{tabular}{c>{\centering\arraybackslash}p{2cm}cccc}
\hline
\textbf{Type} & \textbf{$G_A^{in}$ (Stage 2) in dB} & \multicolumn{3}{c}{\textbf{$G_A^{out}$ (Stage 1) in dB}} \\
\cline{3-5}
 & &  \textbf{-25} & \textbf{-35} & \textbf{-45} \\
\hline
\textbf{RQ4} & \textbf{-8}  & -5.127173 & -0.487680 & 0.677486 \\
  & \textbf{-15}  & -6.918450 & -3.097965 & -0.046018 \\
\hline
\textbf{MQ9}  & \textbf{-8}  & -4.773191 & -0.490571 & 0.704719 \\
              & \textbf{-15} & -7.110577 & -3.121910 & -0.006831 \\
\hline
\end{tabular}
\end{table}

\begin{table}[ht!]
\renewcommand{\arraystretch}{1.3} 
\centering
\caption{$\Delta \text{SNR}$ under varying $G_G^{in}$ Stage 2 during reentry UL.}
\label{tab:reentry-ul-overpowering}
\begin{tabular}{c>{\centering\arraybackslash}p{2cm}c}
\hline
\textbf{Type} & \textbf{$G_G^{in}$ (Stage 2) in dB}& {$\Delta \text{SNR}$ in dB} \\
\hline
\textbf{RQ4} &  \textbf{-15}  & -6.918450  \\
 & \textbf{-10}    & -4.397571  \\
 & \textbf{-5}    & -2.621915 \\
 & \textbf{0}   & -1.149819  \\
\hline
\textbf{MQ9} & \textbf{-15}  & -7.110577  \\
 & \textbf{-10}   & -4.985007   \\
 & \textbf{-5} & -2.221381  \\
 & \textbf{0}  & -1.112713  \\
\hline
\end{tabular}
\end{table}

Fig.~\ref{fig:ul-reentry-power} and Table~\ref{tab:ul-reentry-delta-snr-exp2} illustrate the similarity of UL and DL, where an attacker transmitting at $-25$ dB causes significant signal distortion. Both RQ-4 and MQ-9 exhibit widened power distributions with tails extending toward low-power values, indicating elevated interference.

Table~\ref{tab:reentry-ul-overpowering} further shows that increasing the GSR input gain ($G_G^{in}$) improves $\Delta \text{SNR}$, yet the attacker maintains substantial influence at the $-25$ dB, confirming the UL’s susceptibility under overpowering replay conditions, similar to how we measured for the DL.

\subsection{Launch - DL}
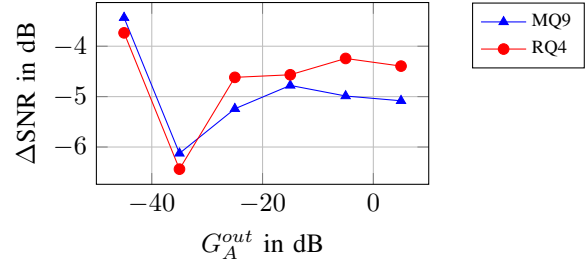
\begin{figure}[tp]
\raggedright
    \begin{tikzpicture}
    \begin{axis}[
        ylabel={$\Delta \text{SNR}$ in dB},
        xlabel={$G_A^{out}$ in dB},
        ylabel near ticks,
        % ymode=log,
        % log basis y=10,
        % ymin=1, ymax=20000,
        % xmin=0, xmax=550,
        % xtick={50, 100, 200, 500},
        % ytick={1, 10, 100, 1000, 10000},
        grid=both,
        height=4cm,
        width=6cm,
        legend style={at={(1.3, 1)}, anchor=north, legend cell align=left, font=\scriptsize}, 
    ]

    \addplot[color=blue, mark=triangle*] coordinates {(-45, -3.433886) (-35, -6.125880) (-25, -5.240992) (-15, -4.777562) (-5, -4.987912) (5, -5.082425)};
    \addlegendentry{MQ9}

    \addplot[color=red, mark=*] coordinates {(-45, -3.734979) (-35, -6.439195) (-25, -4.617735) (-15, -4.566572) (-5, -4.242782) (5, -4.394571)};
    \addlegendentry{RQ4}
    
    % \addplot[color=orange, mark=triangle*] coordinates {(-45, -0.143912) (-35, -0.143912) (-25, -0.143912) (-15, -0.143912) (-5, -0.143912) (5, -0.143912)};
    % \addlegendentry{GS2}
    \end{axis}
\end{tikzpicture}
\caption{$\Delta \text{SNR}$ under varying $G_A^{out}$ during launch DL.}
\label{fig:dl-launch-delta-snr}
\end{figure}

As shown in Fig.~\ref{fig:dl-launch-delta-snr}, the $\Delta \text{SNR}$ remains minimal at low $G_A^{out}$ for both MQ-9 and RQ-4 platforms. Beyond $–45$ dB, the $\Delta \text{SNR}$ begins to degrade more significantly, demonstrating the growing impact of the attacker’s signal on the legitimate communication link. This behavior is similar to the reentry scenario; however, we observe that the attacker requires a lower output gain ($–45$ dB) in the launch DL scenario compared to $–25$ dB during reentry to start overshadowing the legitimate signal, causing destructive interference.

\begin{table}[tp]
\renewcommand{\arraystretch}{1.3} 
\centering
\caption{$\Delta \text{SNR}$ under varying $G_A^{out}$ during launch DL with $G_A^{out}$ set at $-45$ dB and $G_O^{in}$ set at $-15$ dB.}
\label{tab:dl-launch-delta-snr-exp2}
\begin{tabular}{c>{\centering\arraybackslash}p{2cm}ccc}
\hline
\textbf{Type} & \textbf{$G_A^{in}$ (Stage 2) in dB} & \multicolumn{2}{c}{$\Delta \text{SNR}$ in dB} \\
 & &  \textbf{GS1} & \textbf{GS2}\\
\hline
\textbf{RQ4} & \textbf{-8}  & -1.945649  & -0.143912\\
  & \textbf{-15}  & -3.734979 & -0.143912 \\
\hline
\textbf{MQ9}  & \textbf{-8}  & -1.963617 & -0.143912   \\
              & \textbf{-15} & -3.433886 & -0.143912  \\
\hline
\end{tabular}
\end{table}

\begin{table}[ht!]
\renewcommand{\arraystretch}{1.3} 
\centering
\caption{$\Delta \text{SNR}$ under varying $G_O^{in}$ during launch DL.}\label{tab:launch-dl-overpowering}
\begin{tabular}{c>{\centering\arraybackslash}p{1.2cm}ccccc}
\hline
\textbf{Type} & \textbf{ $G_O^{in}$ (Stage 2) in dB} & \multicolumn{2}{c}{\textbf{$G_A^{in}$ (Stage 2) in dB}} & \\
 &  & \multicolumn{2}{c}{\textbf{GS1}} & \textbf{GS2} \\
 \cline{3-4}
 &  & \textbf{-15} & \textbf{-30}&  \\
\hline
\textbf{RQ4} & \textbf{0}  & -6.636374 & -3.952310 & -5.971760 \\
 & \textbf{-5}     & -5.971760 & -3.740749 & -3.952310  \\
 & \textbf{-7}     & -6.304730 & -3.736822 & -2.738036  \\
 & \textbf{-10}    & -6.636374 & -3.580982 & -1.127532 \\
\hline
\textbf{MQ9} & \textbf{0} & -6.934440  & -4.128453 & -5.687190  \\
 & \textbf{-5} & -6.312586 & -4.254511 & -3.921198  \\
 & \textbf{-7} & -7.068297 & -5.764924 & -2.773548  \\
 & \textbf{-10} & -6.597250 & -4.933800 & -1.222542  \\
\hline
\end{tabular}
\end{table}

\begin{table*}[tbp!]
\renewcommand{\arraystretch}{1.2}
\centering
\caption{BER comparison for RQ4 and MQ9 platforms before (Baseline) and after applying the countermeasures.}
\label{tab:combined-ber-results}
\begin{tabular}{c|cc|cc|cc}
\hline
\textbf{Type} 
& \multicolumn{2}{c|}{\textbf{Reentry DL}} 
& \multicolumn{2}{c|}{\textbf{Reentry UL}} 
& \multicolumn{2}{c}{\textbf{Launch DL}} \\
\cline{2-7}
& \textbf{Baseline} & \textbf{Countermeasure} 
& \textbf{Baseline} & \textbf{Countermeasure} 
& \textbf{Baseline} & \textbf{Countermeasure} \\
\hline
\textbf{RQ4} & 10.5995 & 4.8262 & 30.2789 & 3.4504 & 10.1156 & 1.0905 \\
\textbf{MQ9} & 13.0047 & 3.4479 & 26.7524 & 3.4694 & 7.6109 & 1.7057 \\
\hline
\end{tabular}
\end{table*}

Next, by keeping the $G_A^{out}$ fixed at $–45$ dB and varying the $G_A^{in}$, we observe the impact on signal recovery at GS1. As the input gain increases, the $\Delta \text{SNR}$ reported in Table~\ref{tab:dl-launch-delta-snr-exp2} shows a partial recovery of the signal at GS1. However, GS2, unaffected by the attack, maintains a stable $\Delta \text{SNR}$ of approximately $–0.14$ dB. According to the launch scenario specifications, which involve the BFS mechanism \cite{doi:10.2514/6.2014-1689}, GS2 would be consistently chosen as the higher-quality signal source, ensuring that the attack’s impact on the overall communication link is minimized.

Last evaluation is on finding a point of overpowering, we keep the $G_A^{out}$ fixed at $–45$ dB and vary $G_O^{in}$ in Stage 2. The resulting $\Delta \text{SNR}$ values at GS1 and GS2, as shown in Table~\ref{tab:launch-dl-overpowering}. At GS1, as the $G_O^{in}$ increases, we observe some improvement in signal recovery, particularly when $G_O^{in}$ is set to $0$ dB. However, GS2, unaffected by the attack, maintains a relatively stable $\Delta \text{SNR}$ throughout the varying input gains, demonstrating that GS2 is still the more reliable signal source in the presence of interference.

\section{Replay Mitigation Strategy}

% Compliance with security design guidance from \cite{9925759}.

To diminish the impact of overpowering and replay attack, we propose an updated receiver architecture. Before downsampling, the received signal passes through a polyphase matched filter bank improved by the symbol synchronization which is performed by using the timing error detection (TED) method from \cite{walls2017samples}. Moreover, replay attacks introduce phase ambiguities, rendering synchronization and equalization stages that rely on phase information vulnerable to failure. To address this, we replace the traditional constant modulus algorithm (CMA) adaptive filter equalizer with a linear mean-square (LMS) equalizer that operates solely on the DPL output of the symbol synchronizer, thus ignoring phase information. The initialization of CMA assumes a single source and single modulus on the received signal. As $G_A^{out}$ increases, the CMA equalizer begins to converge toward the superposition of both signals. Decision-directed (DD), on the other hand, requires phase coherent symbol detection. Therefore, as long as the received Orion or ground station signal suppresses the attacker, DD convergence is safer. During a replay attack, where the Orion and ground station phases are off, the DD equalization fails, instead of processing the attacker signal. An erroneous signal is desired due to a possible hijacking. In this regard, a more robust CMA is replaced with DD, which is more primitive yet more secure against an extruder attack. Additionally, we tighten the second order loop bandwidth of DPPL for CFO estimation to $15.7$ mrad, allowing faster and more secure phase locking. 

We evaluate this countermeasure under attacker gain thresholds that lead to synchronization failure: $–25$ dB in the reentry scenario and $–45$ dB during launch. Table~\ref{tab:combined-ber-results} reports the BER before and after applying the countermeasure. For the RQ-4, the countermeasure reduces the BER by 54.5\% in the reentry DL, demonstrating a strong improvement in communication quality under attack. For MQ-9, the countermeasure achieves a 73\% reduction in the reentry DL. In the launch scenario, while GS2 (unaffected by the attack) remains the dominant signal source due to the BFS mechanism, the countermeasure applied to GS1 shows a notable 89\% reduction in BER, significantly improving signal recovery at the attacked station. 

\section{Conclusion}
This study demonstrates that RF replay attacks from a HALE-based adversary pose a credible threat to space-ground links during launch and reentry. Using hardware-in-the-loop emulation, we show that both UL and DL channels can be compromised under realistic mission conditions. To mitigate these threats, we propose a secure, energy-efficient receiver design using a phase-coherency-dependent DD equalizer with a narrowed PLL bandwidth. These countermeasures improved synchronization robustness and communication reliability under emulated adversarial conditions. Future work will focus on evaluating the impact of multi-adversary and coordinated attack scenarios where attackers exploit multiple vantage points and combine replay with other attacks to disrupt space-ground links.

% \section{Outline}
% \begin{itemize}
%     \item Introduction: artemis, importance of lunar missions, security emphasis (crewed mission), attack (replay and other kinfs)  and then different type of phases (timeboud characteristics) and say these are the two most vulnerable
%     \item Feasibility/emerging threat landscape/motivation: discuss on HALE capabilities, current physical boundary security (from countermeasures) and the launch/reentry sites
%     \item System model: physical layer, replay attack modeling and the two phases \begin{itemize}
%         \item emphasis on replay attack - how it gets in the system since its geniune 
%     \end{itemize}
%     \item Emulation setup and results: propmsim, sdrs and the post processing-- + results of channels
%     \begin{itemize}
%         \item performance results: diff SNRs and BER
%         \item quantify also probabilities of passed/denied attack
%     \end{itemize}
%     \item Countermeasures
%     \item Conclusion and future direction
%     \item Some notes
%     \begin{itemize}
%         \item use sync and non sync?
%         \item see if we emulate random msgs or genuine msgs 
%         \item energy efficient event driven resilient protocol
%          \item we can decide if we mention the other layers or not -- either here we do it or in the future directions
%     \end{itemize}
% \end{itemize}

% \section{experimental}

% \begin{itemize}
%     \item uplink: gs -> KUS ; attacker at half height and quarter height of SLS in max q
%     \item downlink: gs -> KUS and POD ; attacker is at half height of sls on either side
% \end{itemize}
\bibliographystyle{ieeetr} 
\bibliography{references}

\end{document}